\documentclass[ALICE,manyauthors]{cernphprep}
\usepackage[comma,square,numbers,sort&compress]{natbib}

\usepackage{lineno}
\usepackage{xspace}
\usepackage{hyperref}
\usepackage[T1]{fontenc}
\usepackage{orcidlink}

\def\snn   {\mbox{$\sqrt{s_{_{\rm NN}}}$}\xspace}
\newcommand{\pt}{\ensuremath{p_{\rm T}}\xspace}

\newcommand{\Vn}[1]{\ensuremath{\vec{V}_{#1}}\xspace}

\begin{document}

\begin{titlepage}
\PHyear{2026}       
\PHnumber{014}      
\PHdate{23 Jan}  

\title{Observation of flow vector fluctuations in p--Pb collisions at \mbox{$\mathbf{\sqrt{\textit{s}_{_{\bf NN}}}}=$ 5.02 TeV}}
\ShortTitle{Flow vector fluctuations in p--Pb}   

\Collaboration{ALICE Collaboration\thanks{See Appendix~\ref{app:collab} for the list of collaboration members}}
\ShortAuthor{ALICE Collaboration} 

\begin{abstract}

Measurements of transverse momentum ($\pt$) and pseudorapidity ($\eta$) dependent flow vector fluctuations in p--Pb collisions at $\sqrt{s_{_{\rm NN}}} = 5.02$ TeV at the CERN Large Hadron Collider are presented. By studying long-range two-particle correlations with a template fit method, potential biases from non-flow effects such as jets and resonance decays are effectively suppressed. Significant $\pt$- and $\eta$-dependent fluctuations of the second-harmonic flow vector are observed with more than 5$\sigma$ confidence in p--Pb collisions, similar to the observations in Pb--Pb collisions. The influence of residual non-flow effects has been evaluated and cannot account for the observed fluctuations, thereby confirming the observation of flow vector fluctuations in small collision systems at the LHC. Comparisons to model calculations from 3DGlauber+MUSIC+UrQMD and the parton transport model from AMPT are also presented. The measurements provide constraints on the theoretical modelling of the three-dimensional initial geometry and its event-by-event fluctuations, offering critical insights into the origin of collective flow in small collision systems at the LHC.

\end{abstract}
\end{titlepage}

\setcounter{page}{2} 


\section{Introduction}
\label{sec:intro}

The extensive studies of quark--gluon plasma (QGP) in ultra-relativistic heavy-ion collisions at the Relativistic Heavy Ion Collider (RHIC) and the Large Hadron Collider (LHC) provide unique opportunities to explore the properties of strongly-interacting matter governed by quantum chromodynamics (QCD) under extreme conditions of temperature and energy density\cite{Shuryak:1978ij, Shuryak:1980tp,ALICE:2022wpn,Adcox:2004mh,Arsene:2004fa,Back:2004je,Adams:2005dq,CMS:2024krd}. The anisotropic flow is one of the key phenomena that provides valuable information on the transport properties of the created QGP matter~\cite{Ollitrault:1992bk} and reveals that this matter behaves as a nearly perfect fluid~\cite{Heinz:2013th,Song:2017wtw}. The flow phenomenon characterizes the anisotropic expansion of the produced particles in the final state, which can be quantified by a Fourier decomposition of the single-particle azimuthal distribution~\cite{Voloshin:1994mz},
\begin{linenomath*}
\begin{align}
	\frac{\mathrm{d}^3N}{\mathrm{d}\pt\mathrm{d}\eta\mathrm{d}\varphi}= \frac{\mathrm{d}^2N}{2\pi\mathrm{d}\pt\mathrm{d}\eta}
	\left(1+2\sum_{n=1}^\infty v_{n}(\pt,\eta)\cos[n(\varphi-\Psi_{n})]\right),
    \label{eq:Fourier}
\end{align}
\end{linenomath*}
where $\varphi$ is the azimuthal angle, $\eta$ the pseudorapidity, and \pt the transverse momentum of the produced particles. The $v_{n}(p_{\rm T},\eta)$ and $\Psi_{n}$ are the magnitude and direction of the $n^{\rm th}$-harmonic flow vector $\Vn{n}(\pt,\eta) = v_{n}(\pt,\eta)e^{in\Psi_{n}}$, respectively. The $\Psi_{n}$, also referred to as the flow symmetry plane, is not directly accessible in experiments. Systematic studies on flow and flow fluctuations~\cite{ALICE:2011ab, ATLAS:2012at,Chatrchyan:2013kba,Aad:2013xma,Aad:2014fla, Aad:2015lwa, ALICE:2016kpq,Adam:2016izf,Sirunyan:2017fts,ALICE:2017lyf,Acharya:2017zfg,Acharya:2018lmh,Acharya:2018ihu,Acharya:2019uia, Acharya:2020taj,Acharya:2021hpf,ALICE:2021gxt,CMS:2023bvg,ATLAS:2020sgl} and comprehensive comparisons to calculations using viscous hydrodynamic models~\cite{Heinz:2013th, Luzum:2013yya, Shuryak:2014zxa, Song:2017wtw} have been performed in ultra-relativistic heavy-ion collisions. These studies allow the extraction of the temperature-dependent specific shear and bulk viscosities of the QGP fluid and give unique access to the initial conditions of heavy-ion collisions~\cite{ALICE:2022wpn}. Recently, based on the study of multiparticle azimuthal angle correlations~\cite{Nielsen:2022jms}, flow vector fluctuations have been observed with more than 5$\sigma$ confidence in Pb--Pb collisions at the LHC, which confirms that the flow symmetry plane $\Psi_{n}$ depends on the kinematic regions of $\pt$ and $\eta$~\cite{ALICE:2022dtx,ALICE:2023tvh,ALICE:2024fcv}. These fluctuations directly probe the geometry and density variations inside the colliding nuclei~\cite{Nielsen:2022jms}. These studies enable a more precise understanding of the early stages of heavy-ion collisions in both transverse and longitudinal directions while minimizing the influence of subsequent final-state interactions~\cite{Nielsen:2022jms,Bozek:2018nne,Bozek:2021mov}. This makes them an essential observable for disentangling initial-state effects from those arising during the collision’s dynamic evolution.

Besides the program of colliding heavy ions at ultra-relativistic energies, the LHC also facilitates proton--proton and proton--lead collisions. In these smaller collision systems, the formation of the QGP was not anticipated~\cite{Citron:2018lsq}, and their study was intended as control experiments. Surprisingly, however, novel QCD phenomena have been observed, such as finite anisotropic flow in high-multiplicity events in small collision systems~\cite{ALICE:2013snk, ATLAS:2015hzw, PHENIX:2018lia, CMS:2010ifv,CMS:2014und,CMS:2018loe, ALICE:2019zfl,ALICE:2023gyf,ALICE:2016fzo}, where tens of final-state charged hadrons have been produced. The first such observation came from the discovery of the long-range (in pseudorapidity) two-particle correlations, known as the ``ridge'' structure in p--Pb and pp collisions~\cite{CMS:2010ifv, ALICE:2013snk, ATLAS:2015hzw}. Later, the collective behaviour has been further confirmed in the study of multiparticle azimuthal angle correlations of the produced hadrons~\cite{ALICE:2013snk, ATLAS:2015hzw, PHENIX:2018lia, CMS:2010ifv,CMS:2014und,CMS:2018loe, ALICE:2019zfl,ALICE:2023gyf}. Measurements performed in different collision systems, such as p--Au, d--Au and $^{3}$He--Au at RHIC and the pp, p--Pb and Pb--Pb collisions at the LHC, reveal that the observed anisotropic flow in small collision systems is predominantly driven by the initial geometry and its event-by-event fluctuations~\cite{Song:2017wtw,Grosse-Oetringhaus:2024bwr}. Furthermore, the most recent measurements using identified hadrons have shown that a characteristic mass dependence at low $\pt$, as well as the grouping and splitting of the flow of baryons and mesons at intermediate $\pt$, presenting evidence of partonic collectivity in small collision systems at the LHC~\cite{ALICE:2024vzv}. Most of the measurements mentioned above can be qualitatively or semi-quantitatively described by hydrodynamic calculations~\cite{Bozek:2011if, Mantysaari:2017cni, Weller:2017tsr, Zhao:2017rgg, Zhao:2020pty} or parton transport models~\cite{He:2015hfa, Kurkela:2018ygx}, where the creation of QGP has been implemented. Nevertheless, noticeable discrepancies have been observed between the theoretical model descriptions and the measurements. 
These discrepancies may be due to the incomplete understanding of the spatial structure of the proton, which drives the expansion in such models~\cite{Mantysaari:2022ffw,Zhao:2022ayk}. As confirmed in previous studies~\cite{ALICE:2022dtx,ALICE:2023tvh,ALICE:2024fcv, Nielsen:2022jms}, flow vector fluctuations provide access to information on the geometry and its event-by-event fluctuations in the early stages of collisions. In particular, studying both \pt and $\eta$-dependent flow vector fluctuations could provide valuable constraints on the fluctuating initial geometry in both transverse and longitudinal directions. Thus, it could be an ideal tool to pinpoint existing uncertainties originating from the initial stages in small collision systems. The corresponding measurements in p--Pb collisions and comparisons to theoretical model calculations will be presented in this paper. Note that results aiming at identifying flow vector fluctuations have been reported previously in small collision systems based on the data from the LHC Run 1 program~\cite{ALICE:2017lyf}. There, the two-particle correlations were calculated using the two-particle cumulant method with a pseudorapidity gap of $|\Delta \eta| > $ 1.0. Although significant deviations from unity were observed, the previous analysis was unable to fully disentangle background effects arising from correlations unrelated to the collective expansion of the medium.
These effects are referred to as non-flow~\cite{Heinz:2013th} and are associated with correlations stemming from sources such as the decay of resonances or between particles that originate from the same initial hard-scattering process and subsequently travel in a collimated hadronic structure called a jet.
The new measurements reported in this article improve upon the previous measurement by utilizing the state-of-the-art template fit method, discussed later, and reveal the existence of $\pt$-dependent flow vector fluctuations in p--Pb collisions at the LHC unambiguously.

The structure of this paper is as follows: Section~\ref{sec:method} introduces the observables and the methods applied in the measurements. Section~\ref{sec:exp} describes the experiment setup and the data analysed in this work. The systematic uncertainties are discussed in Section~\ref{sec:statsys}. Finally, the results and summary are presented in Sections~\ref{sec:res} and~\ref{sec:sum}, respectively.

\section{Observables and method}
\label{sec:method}

To study \pt-dependent flow vector fluctuations, the observable $v_2\{{\rm 2PC}\}/v_2[{\rm 2PC}]$, based on two-particle correlations (2PC), is proposed in Ref.~\cite{Heinz:2013bua}. This observable is defined as
\begin{equation}
    \frac{v_{2}\{\rm 2PC\}}{v_{2}[\rm 2PC]}=\frac{V_{2\Delta}(p_{\rm T}^{\rm a},p_{\rm T}^{\rm ref})}{\sqrt{\vphantom{p_{\rm T}^{\rm ref}}V_{2\Delta}(p_{\rm T}^{\rm a},p_{\rm T}^{\rm a}) }\sqrt{V_{2\Delta}(p_{\rm T}^{\rm ref},p_{\rm T}^{\rm ref})}}=\frac{\langle v_{2}(p_{\rm T}^{\rm a}) \, v_{2}(p_{\rm T}^{\rm ref}) \, \cos[2(\psi_{2}(p_{\rm T}^{\rm a})-\psi_{2}(p_{\rm T}^{\rm ref}))]\rangle}{\sqrt{\langle {v_{2}(p_{\rm T}^{\rm a})}^{2}\rangle }\sqrt{\langle {v_{2}(p_{\rm T}^{\rm ref})}^{2}\rangle}}.
    \label{eq:CurlyOverSquare}
\end{equation}
Here, $V_{2\Delta}$ is the two-particle correlation coefficient for the second harmonic, $\pt^{\rm a}$ is taken from a narrow \pt range, while $\pt^{\rm ref}$ is from a wide $\pt$ range. Thus, the difference between $v_2\{\rm 2PC\}$ and $v_2[\rm 2PC]$ is that the former takes the reference flow particles from a wide kinematic range and the particles of interest from a narrow $p_{\rm T}^{\rm a}$ interval, whereas the latter takes both groups of particles from the same narrow $p_{\rm T}^{\rm a}$. This observable quantifies the fluctuations of the flow vector in a narrow \pt interval compared to the flow vector integrated over a wider \pt range, and whether the following relations $\psi_{2}(p_{\rm T}^{\rm a}) = \psi_{2}(p_{\rm T}^{\rm ref})$ or $\langle v_{2}(p_{\rm T}^{\rm a}) \, v_{2}(p_{\rm T}^{\rm ref}) \rangle = \sqrt{\langle {v_{2}(p_{\rm T}^{\rm a})}^{2}\rangle } \, \sqrt{\langle {v_{2}(p_{\rm T}^{\rm ref})}^{2}\rangle}$ hold.
A significantly lower value of $v_2\{{\rm 2PC}\}/v_2[{\rm 2PC}]$ than unity shows the evidence of \pt-dependent flow vector fluctuations.
A similar but more differential observable was also proposed~\cite{Gardim:2012im}, where the flow vector magnitude and symmetry angle between narrow \pt-bins are compared and is defined as
\begin{equation}
    r_{2}(p_{\rm T}^{\rm a},p_{\rm T}^{\rm b})=\frac{V_{2\Delta}(p_{\rm T}^{\rm a},p_{\rm T}^{\rm b})}{\sqrt{V_{2\Delta}(p_{\rm T}^{\rm a},p_{\rm T}^{\rm a})V_{2\Delta}(p_{\rm T}^{\rm b},p_{\rm T}^{\rm b})}}=\frac{\langle v_{2}(p_{\rm T}^{\rm a}) \, v_{2}(p_{\rm T}^{\rm b}) \, \cos[2(\psi_{2}(p_{\rm T}^{\rm a})-\psi_{2}(p_{\rm T}^{\rm b}))]\rangle}{\sqrt{\langle v_{2}(p_{\rm T}^{\rm a})^2\rangle\langle v_{2}(p_{\rm T}^{\rm b})^2\rangle}}.
    \label{eq:rn}
\end{equation}
Different from $v_2\{{\rm 2PC}\}/v_2[{\rm 2PC}]$, here both $\pt^{\rm a}$ and $\pt^{\rm b}$ are taken from narrow \pt intervals. Thus, the $r_{2}(p_{\rm T}^{\rm a},p_{\rm T}^{\rm b})$ observable allows to probe the finer structure of flow vector fluctuations as a function of the kinematic separation $|p_{\rm T}^{\rm a}-p_{\rm T}^{\rm b}|$, although the corresponding statistical uncertainties are larger than those associated with $v_2\{{\rm 2PC}\}/v_2[{\rm 2PC}]$.

At the same time, the $\eta$-dependent flow vector fluctuations have been probed via $r_{2}(\eta^{\rm a},\eta^{\rm b})$, providing evidence of $\eta$-dependent flow vector fluctuations in Pb--Pb collisions at the LHC~\cite{Khachatryan:2015oea,ATLAS:2017rij,ALICE:2023tvh}. These findings reveal challenges in understanding the anisotropic expansion, particularly in the longitudinal direction. 
They help to improve the modelling of longitudinal fluctuations of initial conditions in ultra-relativistic heavy-ion collisions. However, $r_{2}(\eta^{\rm a},\eta^{\rm b})$ is not an ideal observable for asymmetric systems such as p--Pb collisions because $v_{n}(\eta^{\rm a}) \neq v_{n}(-\eta^{\rm a})$. An alternative observable $R_{2}(\eta^{\rm a}, \eta^{\rm b})$ was constructed based on $r_{2}(\eta^{\rm a},\eta^{\rm b})$~\cite{Khachatryan:2015oea}, defined as 
\begin{align}
    &R_{2}(\eta^{\rm a}, \eta^{\rm b}) =\sqrt{r_2(\eta^{\rm a},\eta^{\rm b})r_2(-\eta^{\rm a},-\eta^{\rm b})} =\sqrt{\frac{V_{2\Delta}(\eta^{\rm a},\eta^{\rm b})}{V_{2\Delta}(-\eta^{\rm a},\eta^{\rm b})} \, \frac{V_{2\Delta}(-\eta^{\rm a},-\eta^{\rm b})}{V_{2\Delta}(\eta^{\rm a},-\eta^{\rm b})}} \nonumber \\
&= \sqrt{\frac{\langle v_{2}(\eta^{\rm a})v_{2}(\eta^{\rm b})\, \cos[2(\psi_{2}(\eta^{\rm a})-\psi_{2}(\eta^{\rm b}))]\rangle}{\langle v_{2}(-\eta^{\rm a})v_{2}(\eta^{\rm b}) \, \cos[2(\psi_{2}(-\eta^{\rm a})-\psi_{2}(\eta^{\rm b}))]\rangle} \, \frac{\langle v_{2}(-\eta^{\rm a})v_{2}(-\eta^{\rm b}) \, \cos[2(\psi_{2}(-\eta^{\rm a})-\psi_{2}(-\eta^{\rm b}))]\rangle}{\langle v_{2}(\eta^{\rm a})v_{2}(-\eta^{\rm b})\,\cos[2(\psi_{2}(\eta^{\rm a})-\psi_{2}(-\eta^{\rm b}))]\rangle}}.
\label{eq:r2_eta}
\end{align}
Here, $\eta^{\rm a}$ (or $-\eta^{\rm a}$) is taken from a narrow $\eta$ interval in the positive (or negative) side at midrapidity ($|\eta|<0.8$) and transverse momentum of 0.2$<\pt<$3.0 GeV/$c$, and $\eta^{\rm b}$ (or $-\eta^{\rm b}$) is taken from a different $\eta$ range at forward (or backward) regions. This ensures sufficient separation in pseudorapidity between the two correlated particles, significantly suppressing non-flow contamination in the measurements. Thus, this observable can probe the flow vector fluctuations with different separations in pseudorapidity. 

The observables mentioned above are all constructed based on two-particle correlations, which can be measured using the method used previously in Refs.~\cite{ALICE:2013snk,Khachatryan:2015oea,ALICE:2023gyf}. In this method, the associated yield per trigger particle is constructed as a function of $\Delta\eta$ and $\Delta\varphi$ according to
\begin{equation}
\frac{1}{N_{\mathrm{trig}}}\frac{\mathrm{d}^{2}N_{\mathrm{pair}}}{\mathrm{d}\Delta\eta\mathrm{d}\Delta\varphi} = \frac{S(\Delta\eta,\Delta\varphi)}{B(\Delta\eta,\Delta\varphi)},
\label{eq:C_corr}
\end{equation}
where $N_{\text{trig}}$ is the total number of trigger particles in a given event class and $p_{\rm T}$ interval. 
Here, trigger particles, denoted as $a$, and associated particles, denoted as $b$, are selected from the kinematic regions of $p_{\rm T}^{\rm a}$ (or $\eta_{\rm T}^{\rm a}$) and $p_{\rm T}^{\rm b}$ (or $\eta_{\rm T}^{\rm b}$) defined above, with $\Delta\eta = \eta^{\rm a} - \eta^{\rm b}$ and $\Delta\varphi = \varphi^{\rm a} - \varphi^{\rm b}$.
The $S(\Delta\eta,\Delta\varphi)$ and $B(\Delta\eta,\Delta\varphi)$ are the same and mixed event distributions, the $S(\Delta\eta,\Delta\varphi)$ is constructed by correlating pairs of particles from the same event, normalised by $N_{\text{trig}}$. In contrast, $B(\Delta\eta,\Delta\varphi)$ is constructed by correlating the trigger particles in one event with associated particles from other events in the same multiplicity class within a 2 cm wide interval of primary vertex in the beam direction, $z_{\text{vtx}}$. The associated yield per trigger particle, defined by Eq.~(\ref{eq:C_corr}), is calculated for each $z_{\text{vtx}}$ interval to correct for variations in pair acceptance and efficiency as a function of $z_{\text{vtx}}$. The final correlation function is obtained by averaging the individual ones across the $z_{\text{vtx}}$ intervals, weighted by the number of trigger particles. Finally, the 2D correlation function is projected onto the $\Delta \varphi$-axis ($Y(\Delta \varphi)$). To minimise non-flow contamination from the near-side jet peak in the \pt-dependent study, an additional criterion is applied to the $\eta$ separation between the two correlated particles for the $\Delta \varphi$ projection estimation. Specifically, $|\Delta \eta| > 1.0$ is used for the $v_2\{{\rm 2PC}\}/v_2[{\rm 2PC}]$ measurement and $|\Delta \eta| > 0.8$ for the $r_2$ measurement. This separation in $\eta$ is already satisfied for the $\eta$-dependent observables by correlating particles from midrapidity and forward-backwards rapidity regions.

To further suppress residual non-flow contamination, the template fit method~\cite{ATLAS:2015hzw} is applied. This approach assumes that the functional form of non-flow correlations remains similar across the full centrality or multiplicity range. It disentangles flow-related correlations from non-flow effects by fitting the measured two-particle correlation function with a combination of a baseline (typically derived from peripheral collisions) and a flow-modulated component. Unlike the peripheral subtraction method~\cite{ALICE:2012eyl, CMS:2012qk, ATLAS:2014qaj}, the template fit method allows the baseline to contain a small residual flow signal.
In this analysis, the flow coefficients $V_{n\Delta}$ are obtained by fitting the associated yields per trigger particle from the high-multiplicity event class, denoted $Y^{\rm HM}(\Delta\varphi)$, with a third-order Fourier series, where the first-order term is the non-flow template obtained from the low-multiplicity event class. The expression fitted to the $Y^{\rm HM}(\Delta\varphi)$ is
\begin{equation}
    Y^{\rm HM}(\Delta\varphi)=FY^{\rm LM}(\Delta\varphi)+G\left(1+\sum_{n=2}^{3} V_{n\Delta}\cos(n\Delta\varphi)\right)
    \label{eq:TF}
\end{equation}
$Y^{\rm HM}$ and $Y^{\rm LM}$ are the associated yields per trigger particle in the high-multiplicity and low-multiplicity event classes, respectively. The $F, G, V_{2\Delta}$ and $V_{3\Delta}$ are free parameters for the fit. Meanwhile, $Y^{\rm LM}$ is used as a template for the non-flow estimation. This analysis technique is identical to that used in the previous measurements~\cite{ALICE:2023gyf,ALICE:2024vzv}.

\section{Analysis details}
\label{sec:exp}

The measurements are performed in p--Pb collisions at $\snn=$ 5.02 TeV. The data were collected with ALICE~\cite{ALICE:2008ngc,ALICE:2014sbx} in 2016 during the LHC Run 2 period of data taking. The events considered in this analysis were recorded with a minimum bias trigger using a coincidence signal in the two scintillator arrays of the V0 detector~\cite{Cortese:781854}. They are also used for determining the event multiplicity class. In addition to the minimum bias trigger, it is required that the selected events have a primary vertex within 10 cm of the ALICE nominal interaction point (IP), measured along the beam line. Background events arising from interactions between the beam and residual gas molecules in the beam pipe are removed using information from the V0 and the Silicon Pixel Detector (SPD) detectors. In-bunch pileup is reduced by excluding events with multiple reconstructed vertices. Specific requirements are set for the multiplicity class of the data and template, respectively, with 0--20\% V0A, 20--40\% V0A, and 40--60\% V0A multiplicity classes used for the data and 60--100\% V0A multiplicity class used for the template. In this data sample, approximately 100 million events pass the event selection criteria and are considered in the analysis. 

Concerning the selection criteria of reconstructed charged particles (tracks), the tracks are reconstructed using both the Inner Tracking System (ITS)~\cite{CERN-LHCC-99-012} and the Time Projection Chamber (TPC)~\cite{Dellacasa:451098}. The tracks are required to have at least 70 TPC space points out of a maximum 159 and a reduced $\chi^2$ of the track fit smaller than 4. Tracks are also required to have at least one hit in the SPD, a distance of closest approach to the ALICE IP $<2$ cm in the longitudinal direction (DCA$z$), and a \pt-dependent selection in the transverse direction (DCA${xy}$) ranging from 0.3 cm at 0.2 GeV$/c$ to 0.03 cm at 3.0 GeV$/c$. These criteria lead to negligible contamination from weak decays and background particles emitted when particles interact with the detector material while maintaining a reasonable tracking efficiency, approximately 65\% at $p_{\rm T} =$ 0.2 GeV/$c$ and around 80\% in the higher $p_{\rm T}$ region. 

In addition, analysis-specific selection criteria are applied. When measuring the \pt-dependent observables, tracks are selected at midrapidity, and a minimum separation of one unit in pseudorapidity ($|\Delta \eta| > 1.0$) between correlated tracks is imposed to suppress the correlations from the jet fragments. The \pt range probed in this study is $0.2 <p_{\rm T}< 3.0$ GeV$/c$. By default, forming a two-particle correlation in each kinematic interval (i.e., within a narrow $\pt^{\rm a}$) requires at least two tracks. This ensures that, for example, both $v_{2}\{2\}$ and $v_{2}[2]$ are measured using identical tracks. For the $\eta$-dependent observables, tracks measured by the ITS and the TPC are correlated with hits in the Forward Multiplicity Detector (FMD)~\cite{Cortese:781854}. The FMD detector consists of five different rings of silicon strip detectors covering $1.7 < \eta < 5.0$ (FMDA) and $-3.4 < \eta < -1.70$ (FMDC) and full azimuth~\cite{Cortese:781854}. Due to the FMD's slow data-taking rate, it is necessary to remove the resulting pileup events by eliminating outliers in the correlations between the number of hits measured by the FMD and the multiplicity measured by the V0 detectors, as done in~\cite{ALICE:2023gyf,ALICE:2024vzv}. Correlating the azimuthal angle between particles reconstructed with the ITS and the TPC with those from the FMD hits enables the measurement of long-range correlations. Thus, enforcing an additional $\eta$ separation is unnecessary to suppress the non-flow effects. The number of hits in the FMD is used as a proxy for particle multiplicity. To be able to construct the $\eta$-dependent observable, it is necessary to only use the parts of the FMDA/FMDC with the same $|\eta|$ coverage. Therefore, only hits within $1.8<|\eta_{\rm FMD}|<3.2$ are used.

\section{Systematic uncertainties}
\label{sec:statsys}

The systematic uncertainties of the measurements are evaluated by obtaining the observables with variations in the event or track selection criteria. The requirements on the event primary vertex and FMD pileup are varied to determine the systematic uncertainty of the event selection criterion. For the primary vertex of a given event, it is required that the $z_{\text{vtx}}$ is tightened to be within 7 cm from the ALICE IP instead of 10 cm as the default option. This check yields a negligible effect in the $\pt$-dependent observable but a systematic uncertainty of 1.6\%. A more restrictive requirement on the correlations between multiplicity measured by V0M and FMD is applied, which results in a systematic uncertainty of about 1.5\% for the $\eta$-dependent observable.

Systematic uncertainties arising from the selection criteria imposed at the track level are investigated by changing the track type to include tracks without hits in the SPD. In addition, systematic checks are performed by increasing the minimum number of TPC space points from 70 to 90 and applying more restrictive requirements on the $\chi^2$ per TPC cluster, tightened from 4 to 2.5. None of these checks shows significant deviations from the default analysis. To better understand the detector effects, a Monte Carlo closure test is performed. It compares results obtained at the event generator level from A Multi-Phase Transport (AMPT) model~\cite{Lin:2004en} with the simulation output after the full reconstruction through the GEANT3 simulation~\cite{Brun:1994aa}. The two results are consistent within their uncertainties, and no significant systematic uncertainty was identified in any of the presented measurements. The systematic uncertainty due to limited knowledge on the material budget might be relevant when FMD is involved (i.e., in the pseudorapidity-dependent flow vector fluctuations). The effect is estimated using Monte Carlo simulations with an increased or reduced material budget of the detector descriptions in the GEANT3 simulation by $\pm 10\%$. The resulting effect is found to be negligible.

Finally, potential remaining non-flow contaminations are examined through systematic checks. A minimum $\eta$ separation of $|\eta|>$ 1.0 (0.8) is required for the particle pairs to suppress contributions from the jet peak in the correlation functions for $v_2\{{\rm 2PC}\}/v_2[{\rm 2PC}]$ (for $r_{2}(p_{\rm T}^{\rm a},p_{\rm T}^{\rm b})$). As this criterion is critical for suppressing non-flow correlations, it is checked whether the exact requirement biases the results by enlarging the separation in $\eta$ by 0.2. Note that this check is only necessary for the $p_{\rm T}$-dependent observables. In addition, the residual contributions from non-flow correlations are investigated using predictions from pure non-flow models. DPMJET III~\cite{Roesler:2000he} is used for the $p_{\rm T}$-dependent observable and \textsc{pythia8}~\cite{Sjostrand:2007gs} for the $\eta$-dependent observable. These contributions from remaining non-flow correlations are about 3.5\% for the $v_2\{{\rm 2PC}\}/v_2[{\rm 2PC}]$ and below 13\% for the $r_{2}(p_{\rm T}^{\rm a},p_{\rm T}^{\rm b})$ observable, which are accounted for in the final systematic uncertainties. Meanwhile, the non-flow contributions are negligible for the $\eta$-dependent observable. Only the sources of systematic uncertainty found to be statistically significant, according to the criteria introduced in Ref.~\cite{Barlow:2002yb}, are added in quadrature to obtain the total systematic uncertainty of each observable, which are below 4\% for $v_2\{{\rm 2PC}\}/v_2[{\rm 2PC}]$, less than 13\% for $r_{2}(p_{\rm T}^{\rm a},p_{\rm T}^{\rm b})$ and about 2.2\% for $R_{2}(\eta^{\rm a}, \eta^{\rm b})$.

\section{Results}
\label{sec:res}

\subsection{Transverse momentum dependent flow vector fluctuations}

\begin{figure}[!h]
\centering
\includegraphics[scale=0.6]{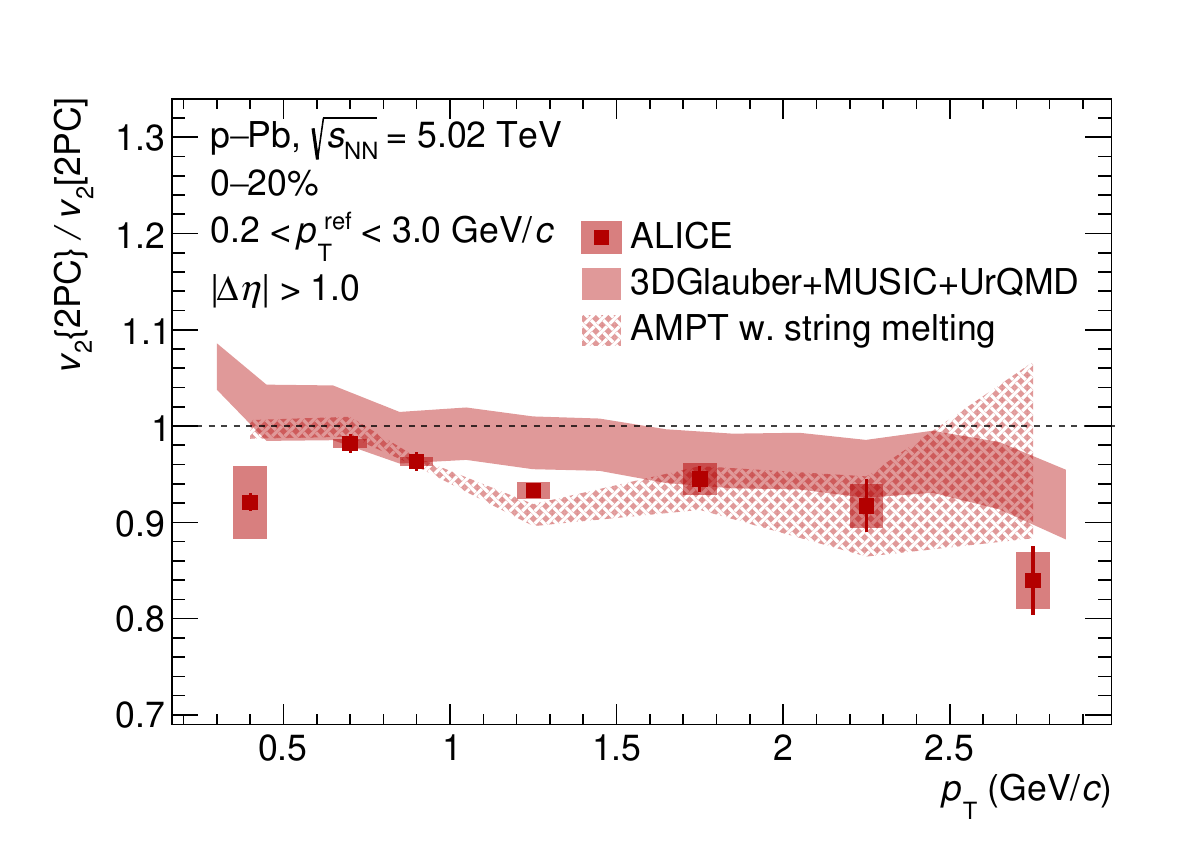}
\caption{The ratio $v_2\{{\rm 2PC}\}/v_2[{\rm 2PC}]$ in p--Pb collisions at $\snn$ = 5.02 TeV as a function of transverse momentum (red points). Statistical (systematic) uncertainties are represented by solid bars (filled boxes). Calculations from AMPT with string melting and 3DGlauber+MUSIC+UrQMD are shown as textured and plain bands, respectively.}
\label{res:v2ratio}
\end{figure}

The results of $v_2\{{\rm 2PC}\}/v_2[{\rm 2PC}]$ are presented as a function of the transverse momentum in Fig~\ref{res:v2ratio}. They were measured in the 0--20\% V0A multiplicity class in p--Pb collisions at $\sqrt{s_{_{\rm NN}}}=5.02$ TeV. The measurements are close to unity for $0.6 < p_{\rm T} < 0.8$ GeV$/c$, where the bulk of the produced particles is located~\cite{ALICE:2014gvd}. 
Excluding the first point, the deviation from unity increases towards higher \pt, exceeding 10\% for $\pt >2.5$ GeV$/c$.
Accounting for both statistical and systematic uncertainties, the measurements for $\pt > 1$ GeV$/c$ deviate from unity with a significance of 7.9$\sigma$, based on the weighted average using the inverse relative uncertainty of the evaluated points. Since potential residual non-flow contributions have been considered in the final systematic uncertainty, the observed deviation from unity cannot be attributed solely to non-flow effects.
This result provides evidence of flow vector fluctuations in p–Pb collisions at the LHC, consistent with observations in Pb–Pb collisions~\cite{ALICE:2022dtx,ALICE:2024fcv}. This similarity might suggest a common driving mechanism behind the observed \pt-dependent flow vector fluctuations in Pb--Pb and p--Pb collisions. It is consistent with the picture established by other collective flow studies in small collision systems~\cite{ALICE:2023gyf,ALICE:2024vzv}.

The flow vector fluctuations are primarily sensitive to the fluctuating initial conditions while being less affected by the complex final-state dynamic evolution~\cite{Nielsen:2022jms}. Comparisons between experimental measurements and theoretical model calculations can provide valuable constraints on the initial conditions, which are poorly known for small collision systems. In this paper, calculations from the AMPT~\cite{Lin:2004en} and 3DGlauber+MUSIC+UrQMD models~\cite{Zhao:2022ayk} are used. These calculations are performed using centrality determinations and particle correlations from the same kinematic regions as those used in the experiment. The AMPT model with string melting is a hybrid model that simulates the dynamics of relativistic heavy-ion collisions at RHIC and the LHC by converting excited strings into partons, allowing for a detailed study of partonic interactions and the subsequent hadronisation process~\cite{Lin:2004en}.
This model successfully reproduced the anisotropic flow measurements in ultra-relativistic heavy-ion collisions~\cite{Zhou:2015eya}. It was also able to qualitatively describe the flow phenomena in small collision systems through a mechanism known as parton escape~\cite{Bzdak:2014dia,He:2015hfa}, based on a small number of parton interactions during the system evolution. The 3DGlauber+MUSIC+UrQMD model is a hybrid approach that combines 3DGlauber initial conditions~\cite{Shen:2017bsr,Shen:2022oyg}, relativistic hydrodynamics with MUSIC~\cite{Schenke:2010rr}, and hadronic transport through UrQMD~\cite{Bass:1998ca,Bleicher:1999xi} to simulate heavy-ion collisions. Its strength lies in its ability to capture both longitudinal and transverse fluctuations in the early stages, providing a more accurate description of the 3D initial geometry. The 3DGlauber+MUSIC+UrQMD model calculations using the kinematics of STAR and PHENIX experiments have partially explained the discrepancy between STAR~\cite{STAR:2022pfn} and PHENIX data~\cite{PHENIX:2018lia} in small system scans, allowing the exploration of the effects of the 3D initial conditions. Unlike the AMPT model, the system created in 3DGlauber+MUSIC+UrQMD goes through a dense partonic rescattering phase (strong-coupling QGP phase). 

\begin{figure}[htbp]
\centering
\includegraphics[scale=0.6]{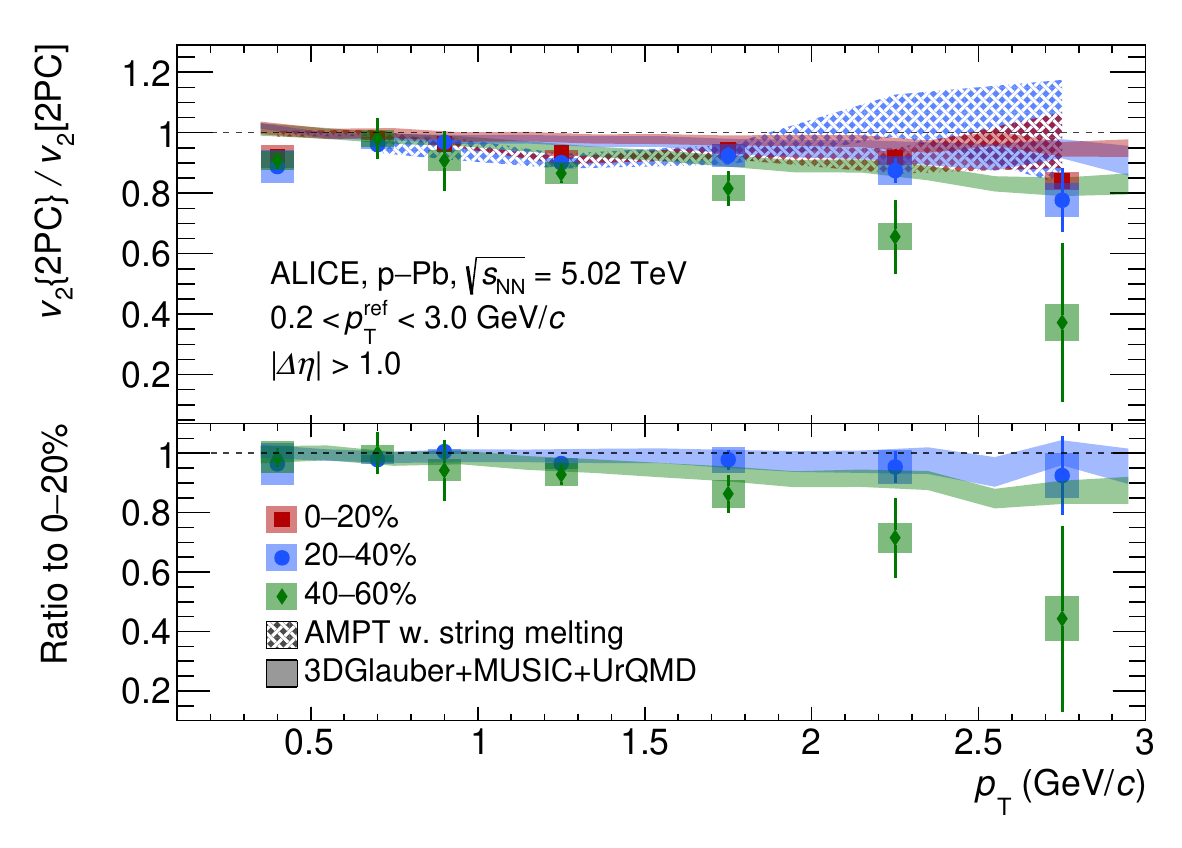}
\caption{The ratio $v_2\{{\rm 2PC}\}/v_2[{\rm 2PC}]$ in p--Pb collisions at $\snn$ = 5.02 TeV as a function of transverse momentum, for 0--20\% V0A (red points), 20--40\% V0A (blue points) and 40--60\% V0A (green points) (top panel). The ratio with respect to the measurements in 0--20\% V0A is also presented (bottom panel). Statistical (systematic) uncertainties are represented by vertical bars (filled boxes). Calculations from AMPT with string melting and 3DGlauber+MUSIC+UrQMD are shown as textured and plain bands, respectively. The ratio results from AMPT are not shown in the bottom panel due to excessively large statistical uncertainties.}
\label{res:v2ratiocompcent}
\end{figure}

Comparisons between the ALICE measurements and calculations from the two models are shown in Fig~\ref{res:v2ratio}. Both models reproduce the general features of the $\pt$-dependent flow vector fluctuations, with an increased deviation from unity towards higher $\pt$. Specifically, the 3DGlauber+MUSIC+UrQMD model predicts a smaller deviation of a few per cent from unity, possibly indicating weaker flow vector fluctuations than those observed in the ALICE measurements. Conversely, the AMPT model may generate slightly stronger flow vector fluctuations and appears to describe the experimental measurements better despite the sizeable statistical uncertainties in the AMPT calculations. 
Moreover, since both models generate final-state anisotropic flow with dominant contributions from partonic interactions, the observed agreements between the ALICE measurements and the model calculations are consistent with the presence of partonic flow in high-multiplicity p--Pb collisions at the LHC.
This aligns with recent observations of partonic flow through measurements of identified hadrons in small collision systems~\cite{ALICE:2024vzv}. Furthermore, $v_2\{{\rm 2PC}\}/v_2[{\rm 2PC}]$ is sensitive to the initial conditions, and not affected by the system's complex dynamic evolution~\cite{Nielsen:2022jms}. Therefore, the presented comparisons provide new constraints on the modelling of the initial geometry and its event-by-event fluctuations in the transverse direction.

\begin{figure}[htbp]
\centering
\includegraphics[scale=0.7]{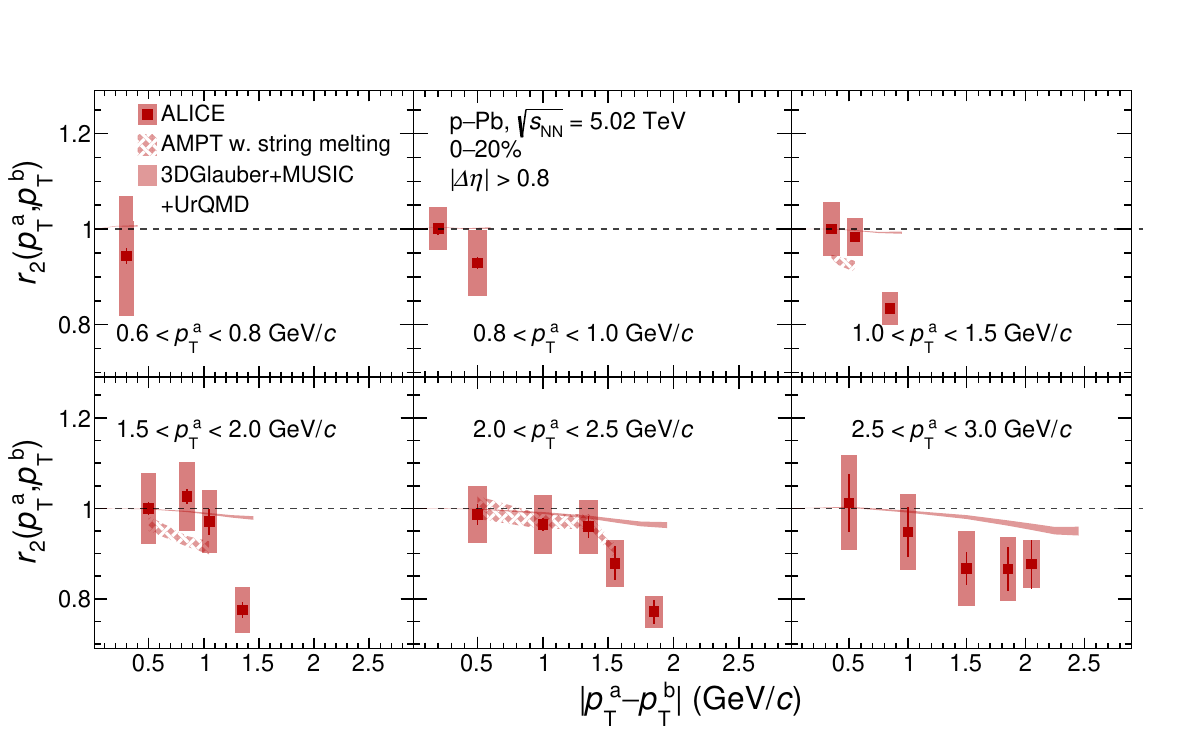}
\caption{Results for $r_2(p_{\rm T}^{\rm a},p_{\rm T}^{\rm b})$ in p--Pb collisions at $\snn$ = 5.02 TeV as a function of transverse momentum (red points). The different panels display results for different $\pt^{\rm a}$ ranges. Statistical (systematic) uncertainties are represented by vertical bars (filled boxes). Calculations from AMPT with string melting and 3DGlauber+MUSIC+UrQMD are shown as textured and plain bands, respectively.}
\label{res:r2}
\end{figure}

The initial geometry might vary significantly in small collision systems depending on the centrality (or multiplicity class)~\cite{Huang:2019tgz}. Understanding how anisotropic flow responds to these changes in the initial geometry and its event-by-event fluctuation in different multiplicity classes allows a better understanding of how the final-state anisotropic flow responds to the size and shape in the initial conditions~\cite{Song:2017wtw}. In addition, the study of the multiplicity dependence of flow observables in small collision systems opens a new window to probe the emergence of collective behaviour and potentially identify the QGP formation in small systems~\cite{Grosse-Oetringhaus:2024bwr}. In Fig.~\ref{res:v2ratiocompcent}, the $v_2\{{\rm 2PC}\}/v_2[{\rm 2PC}]$ was also measured in 20--40\% and 40--60\% V0A multiplicity classes in p--Pb collisions at $\sqrt{s_{_{\rm NN}}}=5.02$ TeV. The trends observed in different multiplicity classes are qualitatively similar. 
For $0.6 < p_{\rm T} < 0.8$ GeV/$c$, the measurements are consistent with unity, followed by a decreasing trend as the $p_{\rm T}$ increases.
The results for 0--20\% and 20--40\% V0A multiplicity classes are compatible within the presented \pt range, while the one in the 40--60\% V0A multiplicity class shows sizeable deviations but with limited significance. Similar to the results presented in Fig.~\ref{res:v2ratio}, the non-flow contaminations in the measurements in 20--40\% and 40--60\% V0A multiplicity classes have been considered in the total systematic uncertainties; such contributions clearly cannot explain the deviations from unity. Thus, the presented results further confirm the existence of flow vector fluctuations not only in the high multiplicity but also in the low multiplicity p--Pb collisions at the LHC. The multiplicity dependence of the flow vector fluctuations, which has a stronger effect in the lower multiplicity region, shows a similar pattern to that observed in peripheral Pb--Pb collisions~\cite{Khachatryan:2015oea}. These results further strengthen the conclusions about a possible common origin of the observed flow vector fluctuations from small to large collision systems.

The measurements are also compared to theoretical model calculations. The 3DGlauber +MUSIC +UrQMD calculation roughly reproduces the multiplicity dependence of the flow vector fluctuations. More specifically, the 3DGlauber+MUSIC+UrQMD model generates a weak deviation from unity for 0--20\% and 20--40\% multiplicity classes. In more peripheral collisions, a more significant deviation, up to 10\%, is observed in the 40--60\% multiplicity class. However, quantitatively speaking, these 3DGlauber+MUSIC +UrQMD calculations tend to slightly underestimate the measured deviation from unity for the last two $\pt$ intervals in the 20--40\% and 40--60\% V0A multiplicity classes. 
The measurements will help to sharpen our quantitative understanding of the initial geometry and the evolution as a function of multiplicity in small collision systems. The AMPT model generates stronger flow vector fluctuations in the 20--40\% multiplicity range than the 3DGlauber+MUSIC+UrQMD calculation, which is compatible with the measurements up to \pt about 1.5 GeV/$c$. However, the sizeable uncertainties of data and model calculations make drawing definitive conclusions difficult in the higher \pt region.

\begin{figure}[htbp]
\centering
\includegraphics[scale=0.7]{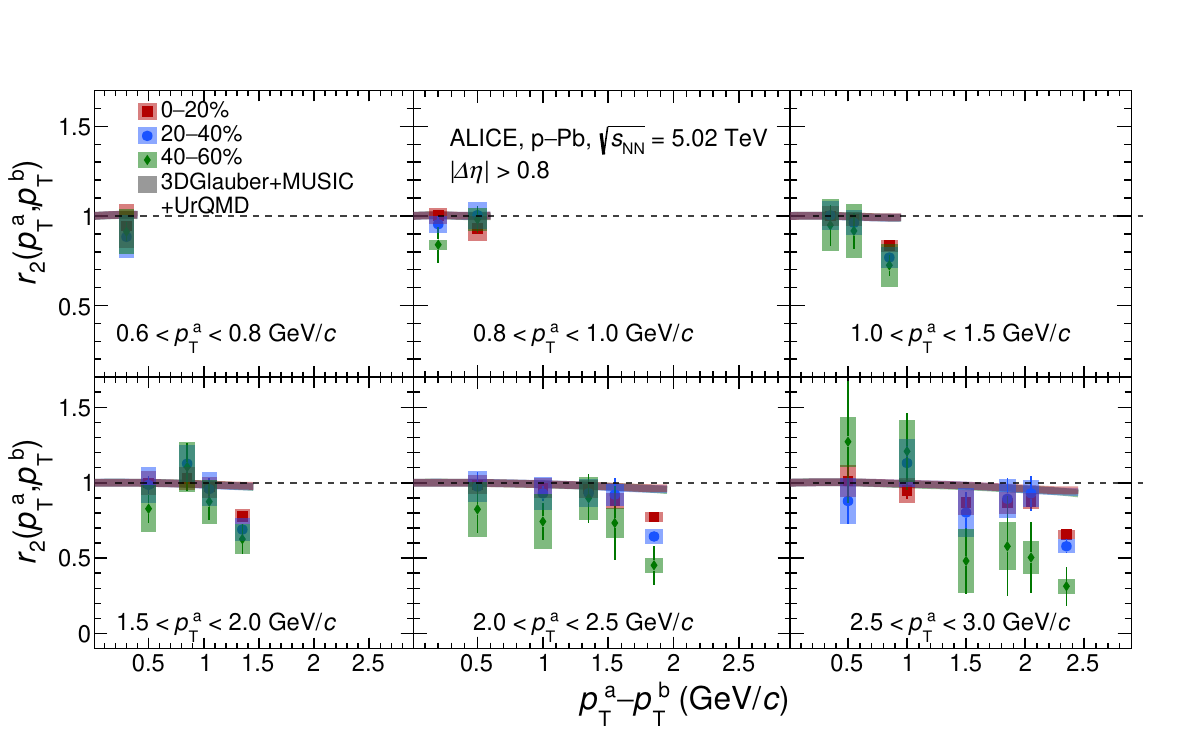}
\caption{Results for $r_2(p_{\rm T}^{\rm a},p_{\rm T}^{\rm b})$ in p--Pb collisions at $\snn$ = 5.02 TeV as a function of transverse momentum for 0--20\% V0A (red points), 20--40\% V0A (blue points) and 40--60\% V0A (green points). Statistical (systematic) uncertainties are represented by vertical bars (filled boxes). Calculations from 3DGlauber+MUSIC+UrQMD are shown as plain bands. The results from AMPT are not shown due to excessively large statistical uncertainties.}
\label{res:r2compcent}
\end{figure}

Besides the study of $v_2\{{\rm 2PC}\}/v_2[{\rm 2PC}]$, the observable $r_2(p_{\rm T}^{\rm a},p_{\rm T}^{\rm b})$ has been measured in 0--20\% V0A p--Pb collisions at $\sqrt{s_{_{\rm NN}}}=5.02$ TeV, to probe the \pt-dependent flow vector fluctuations in narrow \pt intervals. The measurements for different $p_{\rm T}^{\rm a}$ and as a function of $|p_{\rm T}^{\rm a}-p_{\rm T}^{\rm b}|$ are presented in Fig.~\ref{res:r2}. 
For all $p_{\rm T}^{\rm a}$ selections, the measurements show a similar decreasing trend with increasing $|p_{\rm T}^{\rm a}-p_{\rm T}^{\rm b}|$. The deviations from unity start to appear for $|p_{\rm T}^{\rm a}-p_{\rm T}^{\rm b}|>1$ GeV/$c$ and are about 10--20\% for $|p_{\rm T}^{\rm a}-p_{\rm T}^{\rm b}|>1.5$ GeV/$c$. As for $v_2\{{\rm 2PC}\}/v_2[{\rm 2PC}]$, the residual non-flow effect is accounted for in the systematic uncertainties. Thus, the measurements cannot be explained by non-flow correlations but show the presence of \pt-dependent flow vector fluctuations in high multiplicity p--Pb collisions. 
The results are also compared to the theoretical model calculations. On one hand, the AMPT model calculations show compatible results with the measurements, similar to the $v_2\{{\rm 2PC}\}/v_2[{\rm 2PC}]$ results. On the other hand, 3DGlauber+MUSIC+UrQMD calculations slightly underestimate the flow vector fluctuation effect, as one can see in the interval $2.5< p_{\rm T}^{\rm a} < 3.0$ GeV/$c$ shown in the bottom right panel.

The $r_2(p_{\rm T}^{\rm a},p_{\rm T}^{\rm b})$ observable was also measured for the 20--40\% and 40--60\% V0A multiplicity classes. The results are plotted together with the previously discussed results in the 0--20\% class in Fig.~\ref{res:r2compcent}. Generally speaking, the measurements are compatible across centrality classes. The largest deviations of $r_2(p_{\rm T}^{\rm a},p_{\rm T}^{\rm b})$ from unity are again observed for the 20--40\% V0A and 40--60\% V0A multiplicity classes when the $p_{\rm T}^{\rm a}$ is relatively large (see the lower right panel in Fig.~\ref{res:r2compcent}). It is expected that a similar centrality dependence compared to $v_2\{{\rm 2PC}\}/v_2[{\rm 2PC}]$ should be observed in the study of $r_2(p_{\rm T}^{\rm a},p_{\rm T}^{\rm b})$. However, as $r_2(p_{\rm T}^{\rm a},p_{\rm T}^{\rm b})$ probes in more detail the structure of $p_{\rm T}$-dependent flow vector fluctuations, it also requires a larger data sample for accurate measurements. Especially the 40--60\% V0A multiplicity class measurements are affected by low statistical significance. With the data collected through the LHC Run 2 period of data taking in 2016, the measurements suggest a larger deviation of $r_2(p_{\rm T}^{\rm a},p_{\rm T}^{\rm b})$ from unity, however, the relatively large uncertainty on the measurements makes it challenging to achieve a definitive conclusion. The 3DGlauber+MUSIC+UrQMD model calculations may have underestimated the effect of the measured flow vector fluctuations. This is likely true for the 40--60\% multiplicity range, where a much weaker deviation is observed in the 3DGlauber+MUSIC+UrQMD calculations compared to the corresponding measurements.

Overall, the agreement between measurements of \pt-dependent flow vector fluctuations and parton transport model, as well as hydrodynamic model calculations, provides additional information on the origins of flow in small collision systems. It confirms the observation of the anisotropic flow phenomenon in small collision systems from a novel perspective and suggests the presence of partonic flow in p--Pb collisions.

\subsection{Pseudorapidity dependent flow vector fluctuations}

\begin{figure}[htbp]
\centering
\includegraphics[scale=0.6]{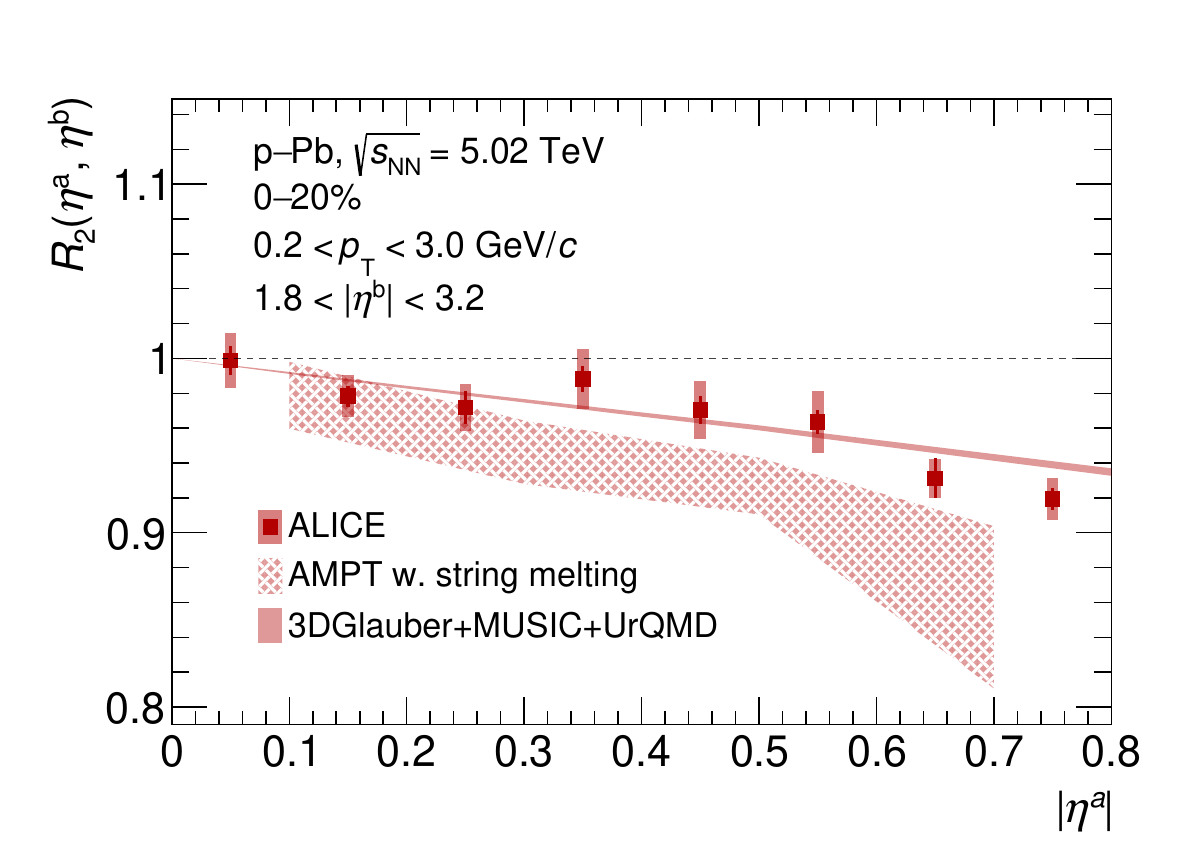}
\caption{Results for $R_{2}(\eta^{\rm a}, \eta^{\rm b})$ in p--Pb collisions at $\snn$ = 5.02 TeV as a function of $|\eta^{\rm a}|$ (red squares). Statistical (systematic) uncertainties are represented by vertical bars (filled boxes). Calculations from AMPT with string melting and 3DGlauber+MUSIC+UrQMD are shown as filled shadows and solid lines.}
\label{res:r2eta}
\end{figure}

Besides the study of $\pt$-dependent flow vector fluctuations, the $\eta$-dependent flow vector fluctuations have also been investigated. The study of anisotropic flow across a large pseudorapidity coverage allows probing the longitudinal structure of the initial state~\cite{ALICE:2023tvh}. Meanwhile, studying $\eta$-dependent flow vector fluctuations will provide valuable information on the three-dimensional initial conditions and their event-by-event fluctuations. Such an improved understanding of the fluctuating initial conditions helps to determine whether the system evolves hydrodynamically, as in heavy-ion collisions, or alternative mechanisms dominate the dynamics in small collision systems. The results for the observable in 0--20\% V0A multiplicity class in p--Pb collisions are presented in Fig.~\ref{res:r2eta} as a function of $|\eta^{\rm a}|$. An overall decreasing trend as a function of $|\eta^{\rm a}|$ is observed, signifying an increase in the flow vector fluctuations. This demonstrates how the flow vectors fluctuate when separated in pseudorapidity by $|\eta^{\rm a}-\eta^{\rm b}|$ or by $|\eta^{\rm a}+\eta^{\rm b}|$. The deviations from unity are significant with 7.2$\sigma$ confidence for $|\eta^{\rm a}|>0.4$ when taking into account statistical and systematic uncertainties. In particular, as the non-flow effect is expected to be fully suppressed after applying the long-range two-particle correlations with the FMD detector and the template fit method, the observed deviation from unity can be regarded as an observation of $\eta$-dependent flow vector fluctuations in p--Pb collisions. Note that the measurement on $R_{2}(\eta^{\rm a}, \eta^{\rm b})$ in p--Pb collisions was first performed by the CMS collaboration~\cite{Khachatryan:2015oea}. It was reported that the deviations from unity were up to $30\%$ for a significantly larger $\eta$. 
However, as discussed above, potential non-flow contaminations might still be present in the early work. These effects have been carefully subtracted in the presented measurements, and any remaining non-flow was found to be negligible based on the DPMJET model study.

\begin{figure}[htbp]
\centering
\includegraphics[scale=0.6]{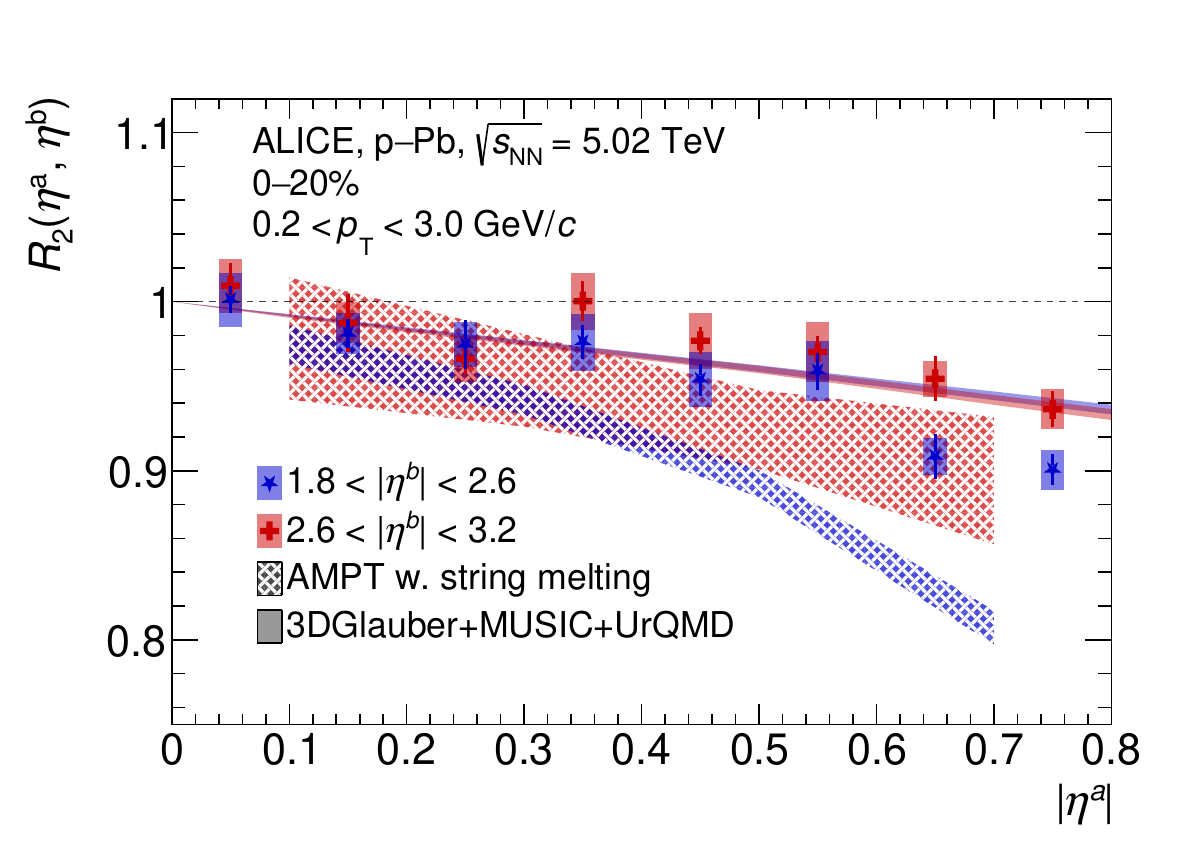}
\caption{Results for $R_{2}(\eta^{\rm a}, \eta^{\rm b})$ in p--Pb collisions at $\snn$ = 5.02 TeV as a function of $|\eta^{\rm a}|$ with $1.8<|\eta^{\rm b}|<2.6$ (blue stars) and $2.6<|\eta^{\rm b}|<3.2$ (red crosses). Statistical (systematic) uncertainties are represented by vertical bars (filled boxes). Calculations from AMPT with string melting and 3DGlauber+MUSIC+UrQMD are shown as textured and plain bands, respectively.}
\label{res:r2eta_split}
\end{figure}

Precise non-flow control also enables meaningful comparison between experimental measurements and the calculations from theoretical models, such as 3DGlauber+MUSIC+UrQMD and AMPT, where the non-flow effects are almost negligible. It is observed in Fig.~\ref{res:r2eta} that both 3DGlauber+MUSIC+UrQMD and the AMPT models generate a clear decreasing trend with increasing $|\eta^{\rm a}|$. More specifically, the 3DGlauber+MUSIC+UrQMD calculations are fully consistent with the measurements, while the AMPT model calculations slightly overestimate the flow vector fluctuations observed in the data. These findings differ somewhat from what was reported in the $p_{\rm T}$-dependent flow vector fluctuations, where 3DGlauber +MUSIC+UrQMD calculations could not quantitatively describe the measurements for the presented \pt intervals. This suggests that the study of flow vector fluctuations in the $p_{\rm T}$ and $\eta$ directions provides independent constraints on the initial conditions of small collision systems. 
Simultaneous descriptions of the measurements of $p_{\rm T}$ and $\eta$-dependent flow vector fluctuations presented in this paper could improve our understanding of anisotropic flow, its fluctuations, and their response to the initial geometry and its three-dimensional variations.

Along with investigating the $|\eta^{\rm a}|$ dependence of $R_{2}(\eta^{\rm a}, \eta^{\rm b})$, an alternative way to probe $\eta$-dependent flow vector fluctuations can be achieved by varying the choices of $|\eta^{\rm b}|$ ranges. This is achieved by selecting $|\eta^{\rm b}|$ from two distinct $\eta$ intervals: $1.8 < |\eta^{\rm b}| < 2.6$ and $2.6 < |\eta^{\rm b}| < 3.2$. The measurements for the 0--20\% V0A multiplicity class are depicted in Fig.~\ref{res:r2eta_split}. In both cases, the results exhibit trends similar to those shown in Fig.~\ref{res:r2eta}, with larger deviations from unity observed in the large $|\eta^{\rm a}|$ region. Additionally, the results for $1.8 < |\eta^{\rm b}| < 2.6$ suggest a larger deviation from unity compared to those for $2.6 < |\eta^{\rm b}| < 3.2$. This observation aligns with expectations, as the relative difference between $|\eta^{\rm a} - \eta^{\rm b}|$ and $|\eta^{\rm a} + \eta^{\rm b}|$ diminishes when $\eta^{\rm b}$ is selected from a pseudorapidity region further away from $\eta^{\rm a}$. Similar findings have recently been reported in Pb--Pb collisions~\cite{ALICE:2023tvh}. Furthermore, the AMPT model calculations successfully reproduce the dependence on the $|\eta^{\rm b}|$ selection despite generating slightly stronger effects of $\eta$-dependent flow vector fluctuations, in particular, the one with $2.6 < |\eta^{\rm b}| < 3.2$. Nevertheless, no difference between the two choices of $|\eta^{\rm b}|$ was found in the 3DGlauber+MUSIC+UrQMD calculations. This model shows compatible results for $2.6 < |\eta^{\rm b}| < 3.2$, but significantly underestimates the effects of $\eta$-dependent flow vector fluctuations in the range $1.8 < |\eta^{\rm b}| < 2.6$, with 8.6$\sigma$ confidence for $|\eta^{\rm a}| > 0.4$.

\begin{figure}[htbp]
\centering
\includegraphics[scale=0.6]{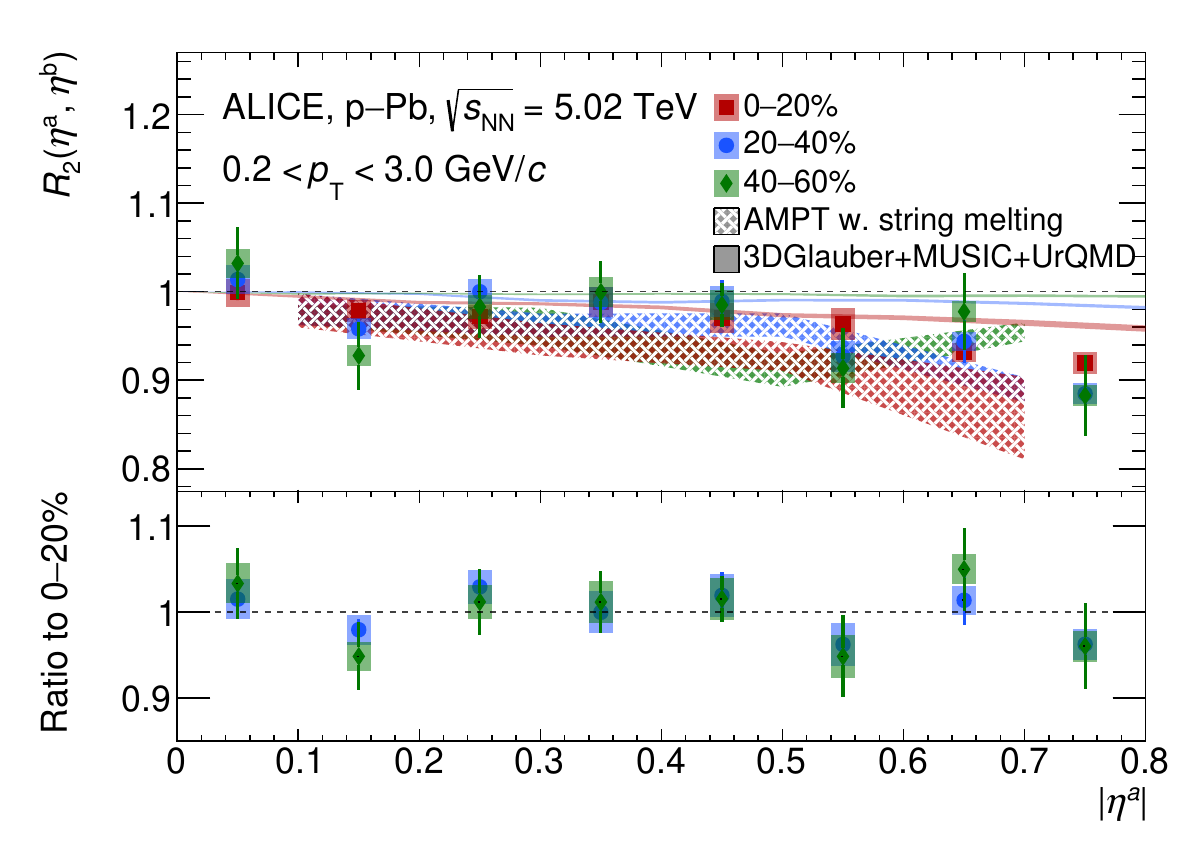}
\caption{Results for $R_{2}(\eta^{\rm a}, \eta^{\rm b})$ in p--Pb collisions at $\snn$ = 5.02 TeV as a function of transverse momentum, with $1.8<\eta^{\rm b}<3.2$ and for V0A 0--20\% (red squares), 20--40\% (blue circles) and 40--60\% (green diamonds). Statistical (systematic) uncertainties are represented by vertical bars (filled boxes). Calculations from AMPT with string melting and 3DGlauber+MUSIC+UrQMD are shown as textured and plain bands, respectively.}
\label{res:r2etacompcent}
\end{figure}

The measurements of $\eta$-dependent flow vector fluctuations were further extended to the 20--40\% and 40--60\% V0A multiplicity classes. The results are presented in Fig.~\ref{res:r2etacompcent}. Unlike the $p_{\rm T}$-dependent results, the measurements of $R_{2}(\eta^{\rm a}, \eta^{\rm b})$ in the 0--20\%, 20--40\%, and 40--60\% classes are consistent within uncertainties. The different multiplicity class dependencies of $p_{\rm T}$ and $\eta$-dependent flow vector fluctuations further confirm the importance of studying both in small collision systems. 
Due to considerable statistical uncertainties in the AMPT calculations, they are roughly compatible with the measurements. Meanwhile, the precise calculations from 3DGlauber+MUSIC+UrQMD show an interesting centrality dependence of $R_{2}(\eta^{\rm a}, \eta^{\rm b})$, with the strongest effects observed in the 0--20\% multiplicity class, and the deviation from unity reduced to less than 1\% in the 40--60\% multiplicity class. This centrality dependence cannot yet be confirmed because of sizable measurement uncertainties. However, there is a hint that the 3DGlauber+MUSIC+UrQMD calculations might underestimate the $\eta$-dependent flow vector fluctuations. 
Considering the results shown in Figs.~\ref{res:r2eta},~\ref{res:r2eta_split},~\ref{res:r2etacompcent}, neither the 3DGlauber+MUSIC+UrQMD nor the AMPT model could quantitatively reproduce the measured $\eta$-dependent flow vector fluctuations. The above comparisons between experimental measurements and theoretical model calculations reveal that while the AMPT model better captures the initial geometric effects in the transverse direction compared to the 3DGlauber+MUSIC+UrQMD model, neither model performs well in describing the fluctuations in the longitudinal directions. It is clear that the presented measurements provide new insight into the three-dimensional initial conditions in a small collision system, which were not well constrained by the existing measurements on $\eta$-dependent particle production and anisotropic flow alone.

\section{Summary}
\label{sec:sum}

The transverse momentum (\pt) and pseudorapidity ($\eta$) dependent flow vector fluctuations have been studied using the observables $v_2\{{\rm 2PC}\}/v_2[{\rm 2PC}]$, $r_2(p_{\rm T}^{\rm a},p_{\rm T}^{\rm b})$ and $R_{2}(\eta^{\rm a}, \eta^{\rm b})$ in $\snn=5.02$ p--Pb collisions. The measurements are performed using long-range two-particle correlations with the state-of-the-art template fit method to suppress non-flow contaminations. Significant deviations from unity with more than 5$\sigma$ confidence are observed for all observables, confirming the presence of \pt and $\eta$-dependent flow vector fluctuations in p--Pb collisions at the LHC. The measurements are also performed in different multiplicity classes. The results of $v_2\{{\rm 2PC}\}/v_2[{\rm 2PC}]$ and $r_2(p_{\rm T}^{\rm a},p_{\rm T}^{\rm b})$ show a smaller deviation from unity in the 0--20\% V0A multiplicity class compared to those from the 40--60\% V0A multiplicity class. These suggest more pronounced \pt-dependent flow vector fluctuations in the low-multiplicity class in p--Pb collisions. Meanwhile, the study of $\eta$-dependent flow vector fluctuations shows no significant difference among different multiplicity classes. The measurements are compared to the calculations from the 3DGlauber+MUSIC+UrQMD model and the AMPT model with string melting. Despite some discrepancies, qualitative agreement between model calculations and experimental measurements is observed. This further supports the evidence for anisotropic flow in small collision systems from a novel perspective and is consistent with the presence of partonic flow in p--Pb collisions.

While theoretical models can reasonably describe the $\pt$ and $\eta$-dependent flow coefficients in small collision systems, the newly accessed flow‑vector fluctuations introduce additional constraints because of their unique sensitivity to the initial geometry and its event‑by‑event variations. This sensitivity provides novel constraints on modelling the initial conditions of small collision systems, extending beyond what flow‑coefficient studies alone can reveal. The corresponding measurements presented in this paper, along with comparisons to theoretical model calculations, provide constraints on the initial conditions in the transverse and longitudinal directions, and can serve as valuable experimental inputs for future 3-dimensional Bayesian analyses. This work paves the way for precisely determining the origins of the anisotropic flow phenomena observed in small collision systems at the LHC~\cite{Citron:2018lsq}.


\newenvironment{acknowledgement}{\relax}{\relax}
\begin{acknowledgement}
\section*{Acknowledgements}
The ALICE Collaboration would like to thank Wenbin Zhao for providing the latest calculations from the state-of-the-art 3DGlauber+MUSIC+UrQMD models.


The ALICE Collaboration would like to thank all its engineers and technicians for their invaluable contributions to the construction of the experiment and the CERN accelerator teams for the outstanding performance of the LHC complex.
The ALICE Collaboration gratefully acknowledges the resources and support provided by all Grid centres and the Worldwide LHC Computing Grid (WLCG) collaboration.
The ALICE Collaboration acknowledges the following funding agencies for their support in building and running the ALICE detector:
A. I. Alikhanyan National Science Laboratory (Yerevan Physics Institute) Foundation (ANSL), State Committee of Science and World Federation of Scientists (WFS), Armenia;
Austrian Academy of Sciences, Austrian Science Fund (FWF): [M 2467-N36] and Nationalstiftung f\"{u}r Forschung, Technologie und Entwicklung, Austria;
Ministry of Communications and High Technologies, National Nuclear Research Center, Azerbaijan;
Rede Nacional de Física de Altas Energias (Renafae), Financiadora de Estudos e Projetos (Finep), Funda\c{c}\~{a}o de Amparo \`{a} Pesquisa do Estado de S\~{a}o Paulo (FAPESP) and The Sao Paulo Research Foundation  (FAPESP), Brazil;
Bulgarian Ministry of Education and Science, within the National Roadmap for Research Infrastructures 2020-2027 (object CERN), Bulgaria;
Ministry of Education of China (MOEC) , Ministry of Science \& Technology of China (MSTC) and National Natural Science Foundation of China (NSFC), China;
Ministry of Science and Education and Croatian Science Foundation, Croatia;
Centro de Aplicaciones Tecnol\'{o}gicas y Desarrollo Nuclear (CEADEN), Cubaenerg\'{\i}a, Cuba;
Ministry of Education, Youth and Sports of the Czech Republic, Czech Republic;
The Danish Council for Independent Research | Natural Sciences, the VILLUM FONDEN and Danish National Research Foundation (DNRF), Denmark;
Helsinki Institute of Physics (HIP), Finland;
Commissariat \`{a} l'Energie Atomique (CEA) and Institut National de Physique Nucl\'{e}aire et de Physique des Particules (IN2P3) and Centre National de la Recherche Scientifique (CNRS), France;
Bundesministerium f\"{u}r Forschung, Technologie und Raumfahrt (BMFTR) and GSI Helmholtzzentrum f\"{u}r Schwerionenforschung GmbH, Germany;
National Research, Development and Innovation Office, Hungary;
Department of Atomic Energy Government of India (DAE), Department of Science and Technology, Government of India (DST), University Grants Commission, Government of India (UGC) and Council of Scientific and Industrial Research (CSIR), India;
National Research and Innovation Agency - BRIN, Indonesia;
Istituto Nazionale di Fisica Nucleare (INFN), Italy;
Japanese Ministry of Education, Culture, Sports, Science and Technology (MEXT) and Japan Society for the Promotion of Science (JSPS) KAKENHI, Japan;
Consejo Nacional de Ciencia (CONACYT) y Tecnolog\'{i}a, through Fondo de Cooperaci\'{o}n Internacional en Ciencia y Tecnolog\'{i}a (FONCICYT) and Direcci\'{o}n General de Asuntos del Personal Academico (DGAPA), Mexico;
Nederlandse Organisatie voor Wetenschappelijk Onderzoek (NWO), Netherlands;
The Research Council of Norway, Norway;
Pontificia Universidad Cat\'{o}lica del Per\'{u}, Peru;
Ministry of Science and Higher Education, National Science Centre and WUT ID-UB, Poland;
Korea Institute of Science and Technology Information and National Research Foundation of Korea (NRF), Republic of Korea;
Ministry of Education and Scientific Research, Institute of Atomic Physics, Ministry of Research and Innovation and Institute of Atomic Physics and Universitatea Nationala de Stiinta si Tehnologie Politehnica Bucuresti, Romania;
Ministerstvo skolstva, vyskumu, vyvoja a mladeze SR, Slovakia;
National Research Foundation of South Africa, South Africa;
Swedish Research Council (VR) and Knut \& Alice Wallenberg Foundation (KAW), Sweden;
European Organization for Nuclear Research, Switzerland;
Suranaree University of Technology (SUT), National Science and Technology Development Agency (NSTDA) and National Science, Research and Innovation Fund (NSRF via PMU-B B05F650021), Thailand;
Turkish Energy, Nuclear and Mineral Research Agency (TENMAK), Turkey;
National Academy of  Sciences of Ukraine, Ukraine;
Science and Technology Facilities Council (STFC), United Kingdom;
National Science Foundation of the United States of America (NSF) and United States Department of Energy, Office of Nuclear Physics (DOE NP), United States of America.
In addition, individual groups or members have received support from:
Czech Science Foundation (grant no. 23-07499S), Czech Republic;
FORTE project, reg.\ no.\ CZ.02.01.01/00/22\_008/0004632, Czech Republic, co-funded by the European Union, Czech Republic;
European Research Council (grant nos. 101220549, 950692), European Union;
Deutsche Forschungs Gemeinschaft (DFG, German Research Foundation) ``Neutrinos and Dark Matter in Astro- and Particle Physics'' (grant no. SFB 1258), Germany;
CONVECS project, CUP C97H23001700002 FESR 2021-2027 program, Italy;
FAIR - Future Artificial Intelligence Research, funded by the NextGenerationEU program (Italy).

\end{acknowledgement}

\bibliographystyle{utphys}   
\bibliography{bibliography}

\newpage
\appendix

%
%

\section{The ALICE Collaboration}
\label{app:collab}
\begin{flushleft} 
\small

D.A.H.~Abdallah\,\orcidlink{0000-0003-4768-2718}\,$^{\rm 134}$, 
I.J.~Abualrob\,\orcidlink{0009-0005-3519-5631}\,$^{\rm 112}$, 
S.~Acharya\,\orcidlink{0000-0002-9213-5329}\,$^{\rm 49}$, 
G.~Aglieri Rinella\,\orcidlink{0000-0002-9611-3696}\,$^{\rm 32}$, 
L.~Aglietta\,\orcidlink{0009-0003-0763-6802}\,$^{\rm 24}$, 
N.~Agrawal\,\orcidlink{0000-0003-0348-9836}\,$^{\rm 25}$, 
Z.~Ahammed\,\orcidlink{0000-0001-5241-7412}\,$^{\rm 132}$, 
S.~Ahmad\,\orcidlink{0000-0003-0497-5705}\,$^{\rm 15}$, 
I.~Ahuja\,\orcidlink{0000-0002-4417-1392}\,$^{\rm 36}$, 
Z.~Akbar$^{\rm 79}$
% ...

\section*{Affiliation Notes}

$^{\rm I}$ Deceased\\
$^{\rm II}$ Also at: Fondazione Bruno Kessler (FBK), Trento, Italy\\
$^{\rm III}$ Also at: Max-Planck-Institut fur Physik, Munich, Germany\\
$^{\rm IV}$ Also at: Czech Technical University in Prague, Prague, Czech Republic\\
$^{\rm V}$ Also at: Instituto de Fisica da Universidade de Sao Paulo\\
$^{\rm VI}$ Also at: Dipartimento DET del Politecnico di Torino, Turin, Italy\\
$^{\rm VII}$ Also at: University College of Dublin, Dublin, Ireland\\
$^{\rm VIII}$ Also at: Department of Applied Physics, Aligarh Muslim University, Aligarh, India\\
$^{\rm IX}$ Also at: Institute of Theoretical Physics, University of Wroclaw, Wroclaw, Poland\\
$^{\rm X}$ Also at: Facultad de Ciencias, Universidad Nacional Aut\'{o}noma de M\'{e}xico, Mexico City, Mexico\\

\section*{Collaboration Institutes}

$^{1}$ A.I. Alikhanyan National Science Laboratory (Yerevan Physics Institute) Foundation, Yerevan, Armenia\\
$^{2}$ AGH University of Krakow, Cracow, Poland\\
$^{3}$ Bogolyubov Institute for Theoretical Physics, National Academy of Sciences of Ukraine, Kyiv, Ukraine\\
$^{4}$ Bose Institute, Department of Physics  and Centre for Astroparticle Physics and Space Science (CAPSS), Kolkata, India\\
$^{5}$ California Polytechnic State University, San Luis Obispo, California, United States\\
$^{6}$ Central China Normal University, Wuhan, China\\
$^{7}$ Centro de Aplicaciones Tecnol\'{o}gicas y Desarrollo Nuclear (CEADEN), Havana, Cuba\\
$^{8}$ Centro de Investigaci\'{o}n y de Estudios Avanzados (CINVESTAV), Mexico City and M\'{e}rida, Mexico\\
$^{9}$ Chicago State University, Chicago, Illinois, United States\\
$^{10}$ China Nuclear Data Center, China Institute of Atomic Energy, Beijing, China\\
$^{11}$ China University of Geosciences, Wuhan, China\\
$^{12}$ Chungbuk National University, Cheongju, Republic of Korea\\
$^{13}$ Comenius University Bratislava, Faculty of Mathematics, Physics and Informatics, Bratislava, Slovak Republic\\
$^{14}$ Creighton University, Omaha, Nebraska, United States\\
$^{15}$ Department of Physics, Aligarh Muslim University, Aligarh, India\\
$^{16}$ Department of Physics, Pusan National University, Pusan, Republic of Korea\\
$^{17}$ Department of Physics, Sejong University, Seoul, Republic of Korea\\
$^{18}$ Department of Physics, University of California, Berkeley, California, United States\\
$^{19}$ Department of Physics, University of Oslo, Oslo, Norway\\
$^{20}$ Department of Physics and Technology, University of Bergen, Bergen, Norway\\
$^{21}$ Dipartimento di Fisica, Universit\`{a} di Pavia, Pavia, Italy\\
$^{22}$ Dipartimento di Fisica dell'Universit\`{a} and Sezione INFN, Cagliari, Italy\\
$^{23}$ Dipartimento di Fisica dell'Universit\`{a} and Sezione INFN, Trieste, Italy\\
$^{24}$ Dipartimento di Fisica dell'Universit\`{a} and Sezione INFN, Turin, Italy\\
$^{25}$ Dipartimento di Fisica e Astronomia dell'Universit\`{a} and Sezione INFN, Bologna, Italy\\
$^{26}$ Dipartimento di Fisica e Astronomia dell'Universit\`{a} and Sezione INFN, Catania, Italy\\
$^{27}$ Dipartimento di Fisica e Astronomia dell'Universit\`{a} and Sezione INFN, Padova, Italy\\
$^{28}$ Dipartimento di Fisica `E.R.~Caianiello' dell'Universit\`{a} and Gruppo Collegato INFN, Salerno, Italy\\
$^{29}$ Dipartimento DISAT del Politecnico and Sezione INFN, Turin, Italy\\
$^{30}$ Dipartimento di Scienze MIFT, Universit\`{a} di Messina, Messina, Italy\\
$^{31}$ Dipartimento Interateneo di Fisica `M.~Merlin' and Sezione INFN, Bari, Italy\\
$^{32}$ European Organization for Nuclear Research (CERN), Geneva, Switzerland\\
$^{33}$ Faculty of Electrical Engineering, Mechanical Engineering and Naval Architecture, University of Split, Split, Croatia\\
$^{34}$ Faculty of Nuclear Sciences and Physical Engineering, Czech Technical University in Prague, Prague, Czech Republic\\
$^{35}$ Faculty of Physics, Sofia University, Sofia, Bulgaria\\
$^{36}$ Faculty of Science, P.J.~\v{S}af\'{a}rik University, Ko\v{s}ice, Slovak Republic\\
$^{37}$ Faculty of Technology, Environmental and Social Sciences, Bergen, Norway\\
$^{38}$ Frankfurt Institute for Advanced Studies, Johann Wolfgang Goethe-Universit\"{a}t Frankfurt, Frankfurt, Germany\\
$^{39}$ Fudan University, Shanghai, China\\
$^{40}$ Gauhati University, Department of Physics, Guwahati, India\\
$^{41}$ Helmholtz-Institut f\"{u}r Strahlen- und Kernphysik, Rheinische Friedrich-Wilhelms-Universit\"{a}t Bonn, Bonn, Germany\\
$^{42}$ Helsinki Institute of Physics (HIP), Helsinki, Finland\\
$^{43}$ High Energy Physics Group,  Universidad Aut\'{o}noma de Puebla, Puebla, Mexico\\
$^{44}$ Horia Hulubei National Institute of Physics and Nuclear Engineering, Bucharest, Romania\\
$^{45}$ HUN-REN Wigner Research Centre for Physics, Budapest, Hungary\\
$^{46}$ Indian Institute of Technology Bombay (IIT), Mumbai, India\\
$^{47}$ Indian Institute of Technology Indore, Indore, India\\
$^{48}$ INFN, Laboratori Nazionali di Frascati, Frascati, Italy\\
$^{49}$ INFN, Sezione di Bari, Bari, Italy\\
$^{50}$ INFN, Sezione di Bologna, Bologna, Italy\\
$^{51}$ INFN, Sezione di Cagliari, Cagliari, Italy\\
$^{52}$ INFN, Sezione di Catania, Catania, Italy\\
$^{53}$ INFN, Sezione di Padova, Padova, Italy\\
$^{54}$ INFN, Sezione di Pavia, Pavia, Italy\\
$^{55}$ INFN, Sezione di Torino, Turin, Italy\\
$^{56}$ INFN, Sezione di Trieste, Trieste, Italy\\
$^{57}$ Inha University, Incheon, Republic of Korea\\
$^{58}$ Institute for Gravitational and Subatomic Physics (GRASP), Utrecht University/Nikhef, Utrecht, Netherlands\\
$^{59}$ Institute of Experimental Physics, Slovak Academy of Sciences, Ko\v{s}ice, Slovak Republic\\
$^{60}$ Institute of Physics, Homi Bhabha National Institute, Bhubaneswar, India\\
$^{61}$ Institute of Physics of the Czech Academy of Sciences, Prague, Czech Republic\\
$^{62}$ Institute of Space Science (ISS), Bucharest, Romania\\
$^{63}$ Institut f\"{u}r Kernphysik, Johann Wolfgang Goethe-Universit\"{a}t Frankfurt, Frankfurt, Germany\\
$^{64}$ Instituto de Ciencias Nucleares, Universidad Nacional Aut\'{o}noma de M\'{e}xico, Mexico City, Mexico\\
$^{65}$ Instituto de F\'{i}sica, Universidade Federal do Rio Grande do Sul (UFRGS), Porto Alegre, Brazil\\
$^{66}$ Instituto de F\'{\i}sica, Universidad Nacional Aut\'{o}noma de M\'{e}xico, Mexico City, Mexico\\
$^{67}$ iThemba LABS, National Research Foundation, Somerset West, South Africa\\
$^{68}$ Jeonbuk National University, Jeonju, Republic of Korea\\
$^{69}$ Korea Institute of Science and Technology Information, Daejeon, Republic of Korea\\
$^{70}$ Laboratoire de Physique Subatomique et de Cosmologie, Universit\'{e} Grenoble-Alpes, CNRS-IN2P3, Grenoble, France\\
$^{71}$ Lawrence Berkeley National Laboratory, Berkeley, California, United States\\
$^{72}$ Lund University Department of Physics, Division of Particle Physics, Lund, Sweden\\
$^{73}$ Marietta Blau Institute, Vienna, Austria\\
$^{74}$ Nagasaki Institute of Applied Science, Nagasaki, Japan\\
$^{75}$ Nara Women{'}s University (NWU), Nara, Japan\\
$^{76}$ National Centre for Nuclear Research, Warsaw, Poland\\
$^{77}$ National Institute of Science Education and Research, Homi Bhabha National Institute, Jatni, India\\
$^{78}$ National Nuclear Research Center, Baku, Azerbaijan\\
$^{79}$ National Research and Innovation Agency - BRIN, Jakarta, Indonesia\\
$^{80}$ Niels Bohr Institute, University of Copenhagen, Copenhagen, Denmark\\
$^{81}$ Nikhef, National institute for subatomic physics, Amsterdam, Netherlands\\
$^{82}$ Nuclear Physics Group, STFC Daresbury Laboratory, Daresbury, United Kingdom\\
$^{83}$ Nuclear Physics Institute of the Czech Academy of Sciences, Husinec-\v{R}e\v{z}, Czech Republic\\
$^{84}$ Oak Ridge National Laboratory, Oak Ridge, Tennessee, United States\\
$^{85}$ Ohio State University, Columbus, Ohio, United States\\
$^{86}$ Physics department, Faculty of science, University of Zagreb, Zagreb, Croatia\\
$^{87}$ Physics Department, Panjab University, Chandigarh, India\\
$^{88}$ Physics Department, University of Jammu, Jammu, India\\
$^{89}$ Physics Program and International Institute for Sustainability with Knotted Chiral Meta Matter (WPI-SKCM$^{2}$), Hiroshima University, Hiroshima, Japan\\
$^{90}$ Physikalisches Institut, Eberhard-Karls-Universit\"{a}t T\"{u}bingen, T\"{u}bingen, Germany\\
$^{91}$ Physikalisches Institut, Ruprecht-Karls-Universit\"{a}t Heidelberg, Heidelberg, Germany\\
$^{92}$ Physik Department, Technische Universit\"{a}t M\"{u}nchen, Munich, Germany\\
$^{93}$ Politecnico di Bari and Sezione INFN, Bari, Italy\\
$^{94}$ Research Division and ExtreMe Matter Institute EMMI, GSI Helmholtzzentrum f\"ur Schwerionenforschung GmbH, Darmstadt, Germany\\
$^{95}$ Saga University, Saga, Japan\\
$^{96}$ Saha Institute of Nuclear Physics, Homi Bhabha National Institute, Kolkata, India\\
$^{97}$ School of Physics and Astronomy, University of Birmingham, Birmingham, United Kingdom\\
$^{98}$ Secci\'{o}n F\'{\i}sica, Departamento de Ciencias, Pontificia Universidad Cat\'{o}lica del Per\'{u}, Lima, Peru\\
$^{99}$ SUBATECH, IMT Atlantique, Nantes Universit\'{e}, CNRS-IN2P3, Nantes, France\\
$^{100}$ Sungkyunkwan University, Suwon City, Republic of Korea\\
$^{101}$ Suranaree University of Technology, Nakhon Ratchasima, Thailand\\
$^{102}$ Technical University of Ko\v{s}ice, Ko\v{s}ice, Slovak Republic\\
$^{103}$ The Henryk Niewodniczanski Institute of Nuclear Physics, Polish Academy of Sciences, Cracow, Poland\\
$^{104}$ The University of Texas at Austin, Austin, Texas, United States\\
$^{105}$ Universidad Aut\'{o}noma de Sinaloa, Culiac\'{a}n, Mexico\\
$^{106}$ Universidade de S\~{a}o Paulo (USP), S\~{a}o Paulo, Brazil\\
$^{107}$ Universidade Estadual de Campinas (UNICAMP), Campinas, Brazil\\
$^{108}$ Universidade Federal do ABC, Santo Andre, Brazil\\
$^{109}$ Universitatea Nationala de Stiinta si Tehnologie Politehnica Bucuresti, Bucharest, Romania\\
$^{110}$ University of Cape Town, Cape Town, South Africa\\
$^{111}$ University of Derby, Derby, United Kingdom\\
$^{112}$ University of Houston, Houston, Texas, United States\\
$^{113}$ University of Jyv\"{a}skyl\"{a}, Jyv\"{a}skyl\"{a}, Finland\\
$^{114}$ University of Kansas, Lawrence, Kansas, United States\\
$^{115}$ University of Liverpool, Liverpool, United Kingdom\\
$^{116}$ University of Science and Technology of China, Hefei, China\\
$^{117}$ University of Silesia in Katowice, Katowice, Poland\\
$^{118}$ University of South-Eastern Norway, Kongsberg, Norway\\
$^{119}$ University of Tennessee, Knoxville, Tennessee, United States\\
$^{120}$ University of the Witwatersrand, Johannesburg, South Africa\\
$^{121}$ University of Tokyo, Tokyo, Japan\\
$^{122}$ University of Tsukuba, Tsukuba, Japan\\
$^{123}$ Universit\"{a}t M\"{u}nster, Institut f\"{u}r Kernphysik, M\"{u}nster, Germany\\
$^{124}$ Universit\'{e} Clermont Auvergne, CNRS/IN2P3, LPC, Clermont-Ferrand, France\\
$^{125}$ Universit\'{e} de Lyon, CNRS/IN2P3, Institut de Physique des 2 Infinis de Lyon, Lyon, France\\
$^{126}$ Universit\'{e} de Strasbourg, CNRS, IPHC UMR 7178, F-67000 Strasbourg, France, Strasbourg, France\\
$^{127}$ Universit\'{e} Paris-Saclay, Centre d'Etudes de Saclay (CEA), IRFU, D\'{e}partment de Physique Nucl\'{e}aire (DPhN), Saclay, France\\
$^{128}$ Universit\'{e}  Paris-Saclay, CNRS/IN2P3, IJCLab, Orsay, France\\
$^{129}$ Universit\`{a} degli Studi di Foggia, Foggia, Italy\\
$^{130}$ Universit\`{a} del Piemonte Orientale, Vercelli, Italy\\
$^{131}$ Universit\`{a} di Brescia, Brescia, Italy\\
$^{132}$ Variable Energy Cyclotron Centre, Homi Bhabha National Institute, Kolkata, India\\
$^{133}$ Warsaw University of Technology, Warsaw, Poland\\
$^{134}$ Wayne State University, Detroit, Michigan, United States\\
$^{135}$ Yale University, New Haven, Connecticut, United States\\
$^{136}$ Yildiz Technical University, Istanbul, Turkey\\
$^{137}$ Yonsei University, Seoul, Republic of Korea\\
$^{138}$ Affiliated with an institute formerly covered by a cooperation agreement with CERN\\
$^{139}$ Affiliated with an international laboratory covered by a cooperation agreement with CERN.\\

\end{flushleft} 
  
\end{document}